\documentclass[conference,a4paper]{IEEEtran}

\usepackage{cite}      
\usepackage[latin1]{inputenc}
\usepackage[T1]{fontenc}
\usepackage{textcomp}
\usepackage{subfigure}
\usepackage{xspace}
\usepackage{ifpdf}
\providecommand{\href}[2]{#2} 
\providecommand{\hypersetup}[1]{}\providecommand{\url}[1]{#1}
\ifpdf 
  \usepackage[pdftex]{graphicx} 
  \usepackage[pdftex,hypertexnames=false,breaklinks=true]{hyperref}
  \pdfcompresslevel=9 \hypersetup{pdfauthor= {Andreas U. Schmidt}}
  \else
  \usepackage[dvips]{graphicx} 
  \usepackage[dvips]{hyperref} 
\fi
\newcommand{\ie}{, {i.e.},\xspace}
\newcommand{\eg}{, {e.g.},\xspace}

\begin{document}

\title{On the Superdistribution of Digital Goods}

\author{\authorblockN{Andreas U.\ Schmidt}
\authorblockA{Fraunhofer Institute for Secure Information Technology SIT\\
Rheinstraße 75, 64295 Darmstadt, Germany\\
Email: andreas.schmidt@sit.fraunhofer.de}}


%

\specialpapernotice{(Invited Paper)}

\maketitle

\begin{abstract}
Business models involving buyers of digital goods in the
distribution process are called superdistribution schemes.
We review the state-of-the art of research and application
of superdistribution and propose systematic approach to market
mechanisms using super-distribution and technical system
architectures supporting it. The limiting conditions on such
markets are of economic, legal, technical, and psychological
nature.

{\copyright 2008 IEEE. Personal use of this material is permitted. However,
permission to reprint/republish this material for advertising or promotional purposes 
or for creating new collective works for resale or redistribution to servers or
lists, or to reuse any copyrighted component of this work in other works must
be obtained from the IEEE.}
\end{abstract}


%
\IEEEpeerreviewmaketitle
\section{Introduction}\label{sec:introduction}
Information systems in general and the distribution of digital content in particular are
dominated by centralised structures rooted in client-server models, and
large efforts have been made for the vertical integration of content 
production, ingestion, and distribution~\cite{AXMEDIS}.
The final transportation of content to the head-ends is nowadays either
digital broadcast\eg DVB~\cite{Reimers2006}, multicast, as for instance envisioned in
3GPP Long-Term Evolution~\cite{3GPPMBMS2008}, or content push~\cite{OMApush2006}.

Peer-to-peer (p2p) systems on the other hand realise a completely different paradigm
for data transport in networks, namely distribution from nodes to other nodes
with little involvement of central instances~\cite{Androutsellis-Theotokis2004}.
File-sharing networks like KaZaA or Gnutella embody this paradigm on the application level,
implementing overlay networks in which users actively (with varied degrees of automation)
re- or \textit{superdistribute} content, in the form of digital files, to other users.

The term superdistribution may have been coined in~\cite{Mori1990,Mori1997}, in any case it hase been around
in information and communication research for some time.
Though the concept lay dormant for quite a while --- perhaps due to the association
with the dominant use of p2p and file sharing by free riders and the copyright wars ---
interest in superdistribution has been rekindled recently in the content producing industry.
The combined size of the most important existing businesses based on content superdistribution 
schemes are of a small scale in comparison to the turnovers of the media industry as a whole.
Nevertheless they prove that the industry is seriously experimenting with the concept.
Most importantly, superdistribution has even been cast in the form of a  standard
for the mobile domain by the Open Mobile Alliance (OMA).

Technically, superdistribution has hitherto been viewed just as a variant of 
Digital Rights Management (DRM)~\cite{DRM, May 2007}, or of p2p systems, and research on its fundamentals is still scarce.
For instance, basic economic questions pertaining to the viability of superdistribution in particular
in competition with free riders have only been examined in our previous work~\cite{AUS_HICSS2008}.
The present paper presents  a first contribution to a treatment of other characteristic 
issues of superdistribution systems, viewed as information systems in their application and economic context.
A system model for generic superdistribution is proposed in Section~\ref{sec:gener-struct-superd}, while
Section~\ref{sec:two-examples} presents two concrete realisations with distinct traits.
Section~\ref{sec:cond-viable-superd} is the core part of the paper which collects the (in our view)
most important research topics on superdistribution and tries to give an overview over the current
state of knowledge in the area. 
The overarching theme here is the technological and economic viability of such systems. 
Section~\ref{sec:some-refl-copyr} tries to put superdistribution into the context of current
socio-economic developments surrounding content distribution, copyright protection, and piracy,
in front of their historic background.
We conclude in Section~\ref{sec:conclusion}.
\section{The general structure of superdistribution 
networks}\label{sec:gener-struct-superd}
Superdistribution is the combined distribution and market scheme for digital
goods involving buyers in the distribution process in such a way that they redistribute
the good to other legitimate buyers.
Here a \textit{digital good} is an information good in the economical sense~\cite{b:SV99,STEG04}, 
which is represented in digital form, regardless of being embodied physically or only 
in intangible form (some use the term
\textit{virtual goods}, coined by~\cite{AH03} and used for 
for information goods in intangible, digital form, and  distributed via electronic networks).
In an active sense, \textit{to superdistribute} means the combined transaction of acquiring 
a good and its (offering for) re-distribution, or resale, and actually transferring it to another node.

Here we argue that existing system models for DRM are to narrow to accommodate for the 
specific features and structures of superdistribution.
In fact, extending DRM into various directions is a recent research trend,
which is triggered by the manifold ways in which users operate with digital goods
for instance in social networks.
For instance, the authors of~\cite{Stini2006} transcend DRM by envisioning a system
in which only the information on ``who owns this digital good'' is managed and thus
agents in the economic network can be given ample freedom\eg to superdistribute it.
In this section, we present a similar approach to extend information management
systems in an appropriate way for superdistribution.
\subsection{Superdistribution networks}\label{sec:superd-netw}
A superdistribution network in the most general sense has two sides.
The first one is the network over which the good is distributed, economically 
a logistics network for the final distribution of the good to the consumer.
If it is an electronic network, it is a particular kind of a content distribution network,
like Akamai, Amazon S3, Corel, CDNetworks, etc.
Whether a good is distributed by ordinary mail, over an electronic network, or
by short-range communication between mobile devices, is immaterial for the classification
as a superdistribution network.
Paradigm examples exist for all three variants: The classic chain latter in the first,
peer-to-peer networks in the second, and superdistribution between mobile devices as
for instance standardised by OMA in the last case.
In all cases, superdistribution is an overlay over an (often general-purpose)
communication or transportation network, like ordinary mail, the Internet, or Bluetooth
ad hoc  communication between mobile devices.
We call this side of a superdistribution network the \textit{content distribution overlay (CDO)}.
The CDO is a directed graph, which in most (reasonable) cases may be assumed to be 
a connected tree.
The CDO graph can be coloured\ie various attributes may be attached to the edges,
a particular example being that the quality of the good may change\eg improved
by a superdistributing node to compete with other resellers.
\begin{figure}[h]
\centerline{\resizebox{0.44\textwidth}{!}{\includegraphics{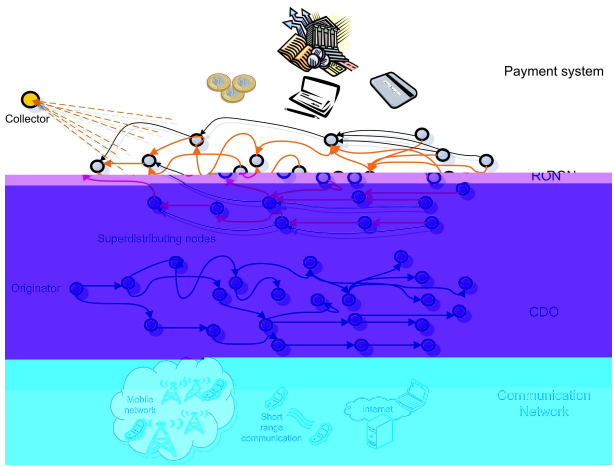}}} 
\caption{Superdistribution overlay networks in the system context.}
\label{fig:sdol}
\end{figure}

Superdistributing nodes in a CDO need a good, economic reason to participate.
This is always true due to the minimum marginal cost greater than zero
incurred by a superdistributing node for storage and transferal of the
good to and from him-/herself (one of the two at least is borne by a specific node).
That is, nodes expect some kind of remuneration for participating
actively in the CDO --- otherwise they may just
become sinks for the digital good.
The flow of remuneration --- pecuniary, informational, immaterial, or of any other 
conceivable kind --- constitutes another overlay network, the 
\textit{remuneration overlay network (RON)}.
The claim here is that no superdistribution network exists without RON, the
most trivial example being the tree spanned by the resale prices paid 
by buyers to superdistributing nodes in the CDO.
In this case the edges of the RON are just the edges of the CDO with
inverted directions (and different colours\eg the sales price, attached). 
The node-set of the RON can be assumed to be a subset of the node-set of the CDO, 
but the relation of the RON's edges to the edges of the CDO is generally nontrivial.
For instance in multi-level marketing, a buyer of a good might pay a reward to  resellers further down the line,
and not only the resales price to his direct reseller. 
Figure~\ref{fig:sdol} shows CDO and RON in the context of underlying communication and payment networks. 
\subsection{Digital goods}\label{sec:digital-goods}
The term digital good used so far refers to the economical atom distributed over
the CDO and being the root cause for the RON. 
Informationally, the digital good is a compound minimally consisting of three
components.

The \textit{content} is the piece of digital information that is actually used
and, if the node chooses to do so, offered for distribution to others.
As the superdistribution network is an economic market mechanism, the content is 
necessarily accompanied by information representing the contractual rules
of a) the global superdistribution market, and b) the particular relationship
between superdistributing (reseller) and acquiring (buyer) node.
Though we will not make use of this distinction of local and global
contract, this orthogonal categorisation may be useful\eg to classify
superdistribution networks.

Using the good means, on the one hand, that the content is
\textit{consumed} by a node who acquires it.
Consumption of the content represents one part of the value proposition that the digital good 
represents to a buyer.
It is governed by a piece of information commonly called the
\textit{consumption licence}, which describes the conditions and permissions under
which the buyer can use the content.
Economically speaking, the consumption licence prescribes the ways in which a
buyer may turn the value proposition of the content into utility.
The consumption licence is also thought to be the informational link between the digital
good and the remuneration overlay by stating the rules of payment for the good
to the superdistributing node, as well as any other reward to be paid to further
nodes or entities.
In this way the consumption licence generates the RON from the CDO, assuming the rules
are adhered to by all participants.
Summarising, the consumption licence consists of three parts:
\begin{itemize}
\item\textit{Consumption rules} describe how content may be used;
\item\textit{Remuneration rules} describe how and who must be paid for it;
\item The \textit{Content association} describes to which content the rules apply.
\end{itemize}

\begin{figure}[h]
\centerline{\resizebox{0.28\textwidth}{!}{\includegraphics{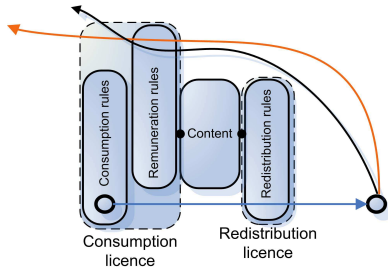}}} 
\caption{Information model of superdistribution.}
\label{fig:licences}
\end{figure}

The second way in which an acquiring node can make use of the good is by superdistribution.
We think of it as governed by rules incorporated in a second licence, the
\textit{redistribution licence}.
Just as the consumption licence connects the good to the RON, the redistribution licence
conditions, or generates the content distribution overlay.
Thus, this licence consists of two essential parts:
\begin{itemize}
\item\textit{Redistribution rules} describe how, to whom and under which conditions content may be redistributed;
\item The \textit{Content association} describes to which content the rules apply.
\end{itemize}
The complete informational structure and its relation to CDO and RON is visualised in
Figure~\ref{fig:licences}

Of course, many other groupings of the information characterising a superdistribution
network are possible --- 
the approach chosen here is lead by the distinction between CDO and RON.
It should also be noted that all notions introduced above are understood here in the
broadest possible sense.
That is large parts of the rules and licences may be represented differently than in
digital form and may include for instance general legislation, copyright law,
social norms, etc.
Redistribution in particular can also be governed by technical conditions\eg
the information system that represents the platform for the execution
of superdistribution.

Thus the particular rules that need to be represented digitally in a concrete 
superdistribution network may be restrictions as well as extensions of 
such global, or external, rules.
Likewise, content associations may be simple titles, digital identifiers denoting
a single piece of content or a group, or be augmented by information protecting
the integrity of digital content such as hash values or signatures.
Nothing restricts the methods by which the licences and the content are
generated, stored, and transferred in the superdistribution network.
This conceptual approach is well known from  general DRM~\cite{DRM}.
\subsection{Examples}\label{sec:examples}
Some more concrete examples might elucidate the abstract notions of Section~\ref{sec:digital-goods}.
The most direct form of remuneration is a resale price paid by the acquiring node to
the superdistributing node.
This makes the superdistribution network a genuine \textit{network market} of buyers/resellers,
where an incentive to buy a good accrues to them by the resale revenues they can achieve.
The ``multi'' in the term \textit{multi-level marketing} often refers to the fact that 
many subsequent levels or generations
of buyers contribute to a node's resales revenues, or even all of them.
This kind of payments or remunerations from the down-line may be restricted
to a finite number of buyer generations or not, the latter case being realised 
in some network marketing schemes for physical goods.

The remunerations may be conditioned by various global or individual factors such
as time, buyer/reseller location, distance in a social network,
or externalities like a measured popularity of the content.
In many cases it makes sense to let a part of the resales price accrue to
a central entity external to the CDO proper, which we call \textit{collector}.
Its role may be to skim revenues from the market for\eg the artists and or labels,
or it may act as a (state) collecting agency implementing a taxation on the
distribution of digital goods.
Examples for second-level payments and the role of the collector are shown
in Figure~\ref{fig:sdol}.

An interesting example for restrictions on the redistribution is the implementation
of territorial protection of some sort.
This can be used to protect resellers from the competition of their (direct)
buyers to a certain extent buy stating\eg ``do not superdistribute before
moving away by 100 metres''.
Thus, this kind of redistribution rule using restrictions based on geographical location
may make particular sense for CDO based
short-range communication between mobile users\ie mobile superdistribution.
We showed in~\cite{KuntzeSchmidtAbendroth2008} how such conditions can be enforced in an efficient, de-central, yet secure manner.
\section{Some examples}\label{sec:two-examples}
As said, superdistribution networks occupy only a small niche even of the online content distribution market.
The better known examples are the following.
Snocap~\cite{SNOCAP}, founded by one of the fathers of Napster,
was started with the idea to obtain licences from the music industry
which explicitly allow to distribute content over the existing, popular p2p
networks. 
Snocap uses audio fingerprinting to track the distribution of content,
and file-sharing networks need to be adapted to support Snocap's remuneration scheme.
Though SnoCap has made some deals with many, even major, labels, it never
took off economically and the company has been aqcuired in February 2008 by the 
social networking platform imeem~\cite{imeem}.
After restructuring and changing the strategy, Snocap has become a general service provider
for online music distribution and for instance provides the technology for 
the music stores in MySpace.
MashBoxx~\cite{mashboxx} started with similar ambitions and also close to the
circles of Napster and Grokster.
The company seems lay dormant for some time, appearing in the news only for
recent intellectual property litigations~\cite{mashboxx-litigation}.
Peer Impact is a pay-for-download file-sharing service created by Wurld Media,
and now acquired together with its parent company 
by the online video service provider Roo~\cite{piacq}.
The file-sharing client has now been re-released under the new brand name ToPeer~\cite{2peer}
, which seems to use part of the original technology to allow p2p users to create
private spaces in which to share content with peers they trust.
In the following we describe two particular examples in more detail.
They are chosen because they represent true technical superdistribution systems, and
 their systems are better documented. 
The two systems represent to some extent oppostite extremes for superdistribution
with respect to (de-)centralisation.
\subsection{Potato System}\label{sec:potato-system}
The Potato system~\cite{potato} is a product developed by the 4FO AG~\cite{4fo} (founded in 2000) 
together with the Fraunhofer Institute for 
Digital Media Technology IDMT~\cite{idmt} in Ilmenau, Germany, for superdistribution of music as mp3-files.
The technical platform for superdistribution presented by the Potato system is centralised, insofar 
as it uses a central accounting service (AS) for registration and publishing new songs by originators,
and to operate the remuneration scheme.
The content CDO is completely free of any DRM measure, the only information protected by the AS (besides the content integrity of which is proved by a hash value) is the redistribution licence,
which is obtained by a buyer upon payment in the form of a transaction number (TAN).
The TAN serves as a receipt which is simply added to the file name, which is in turn announced in
subsequent resales to the AS which initiates the rewarding of resellers.
Some details are found in~\cite{GNACM02, GNWD03}.
Potato supports various payment providers from which the originators of a good may choose.

\begin{figure}[h]
\centerline{\resizebox{0.3\textwidth}{!}{\includegraphics{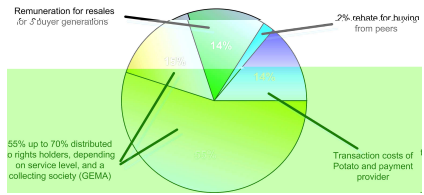}}} 
\caption{Revenue sharing in the Potato system.}
\label{fig:Potato_shares}
\end{figure}
\begin{figure}[h]
\centerline{\resizebox{0.18\textwidth}{!}{\includegraphics{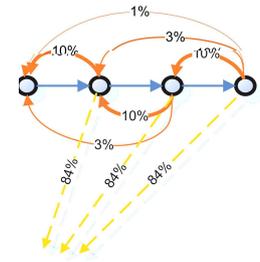}}} 
\caption{CDO and RON of Potato connect 4 buyer generations.}
\label{fig:Potato_local_network}
\end{figure}

The market mechanism and remuneration scheme implemented in the Potato system is perhaps the
most evolved in superdistribution.
The sharing of revenues is shown in Figure~\ref{fig:Potato_shares}.
Potato targets small labels and independent artists, who
 may obtain 55--70\% of the purchase price of every resale, depending on
the service level they choose.
An interesting detail is that Potato has an agreement with the German collecting
society for music, GEMA which obtains the due contributions directly from the system.
Potato itself and the payment provider share 14\% of the purchase price and 
further 14\% are distributed as resale revenues from the buyer to resellers
(this share has been decreased from 35\% in ``Version 1.0'' to the current
``Version 2.0'' value).
The special kind of remuneration for resellers in this system establishes a true multi-level
market with three rewarding levels, each being awarded a geometrically decreasing
share of 10, 3, and 1\%, respectively, cf.~\cite[Section~II.B]{AUS_HICSS2008}.
Thus the CDO and RON look locally as shown in Figure~\ref{fig:Potato_local_network} 
It is interesting to note that a rebate of 2\% (borne by the system, not the resellers) 
is offered for nodes who choose to buy from a peer rather than the central service. 
This an important incentive that reduces the dominant role of a single market
participant, cf.\ Section~\ref{sec:economical-axis}\footnote{The present 
author takes the liberty to claim certain credit for suggesting some of these novel concepts.}.

Originally resellers were mostly left to their own devices in marketing songs for resale.
They could use a resale link containing their TAN on their Web-site or in e-mails.
The most recent developments of the Potato as a superdistribution platform regard 
capabilities to support users in marketing goods\ie means to offer them successfully 
online for superdistribution, and to compete with other resellers.
This includes the extension of resale links to Widgets embodying small online shops where
resellers can display their favourites, covers, and let peers listen to clippings of songs.
4FO also added a social commerce platform SpreadBox~\cite{spreadbox} to its portfolio which also tries to
leverage community aspects of marketing in the form of product recommendation.
\subsection{DRM Paradiso}\label{sec:drm-paradiso}
The Paradiso system~\cite{paradiso}
is a technological solution to DRM-based superdistribution proposed
by Nair Srijith from the research group of Andrew Tannenbaum~\cite{paradiso_IEEE_C}, and others. 
Its central technical trait is that it relies on a \textit{trusted platform}~\cite{FHKS07} to ensure
adherence to consumption and redistribution licence.
Thus it makes some requirements on \textit{compliant devices} with regard to cryptographic
capabilities (hash, AES engine, and PKI management), secure storage, and secure content decoder.

In the content distribution scheme of Paradiso, consumption and remuneration licences are
cryptographically bound to the content and \textit{chained}.
That is, a buyer receives with the content a signed container from the reseller, containing
all previous licences created in every resale upstream in the CDO.
The signature also associates this data to the content.
This enables him\eg to verify that the content has not been tampered with, for instance it 
prevents content masquerading attacks by which a reseller  might try to superdistribute
content of lower quality.
The compliant device can also check that all licence rules have been enforced in all 
previous distribution steps, and enforces the applicable rules for itself\eg respects
and updates the allowed number of resales.
Payment is and out-of band process in Paradiso which is based on a receipt the acquiring node
sends back to the superdistributing node.
It is not hard to see that this system has strong security with respect to the maintenance
of DRM of the content as it is distributed down the CDO. Formal security proofs are given in~\cite{Jonker2006}.

This system provides the strongest possible DRM enforcement in superdistribution which
can be implemented in a completely decentralised fashion.
The system has not been deployed commercially, yet
a prototype has been shown on a Neuros
development board~\cite{neuros}
with similar capabilities of what is typical for
mobile music players (TI 200-MHz ARM926, 120-MHz C54x DSP processor,
64 Mbytes of SDRAM and 10/100 Mbps Ethernet port, etc.). 
The board runs a modified version of the
Linux 2.6 kernel. Based on OpenSSL libraries for cryptographic
support, the higher cryptographic and security functions, are 
emulated by a software layer.
\section{Conditions for viable superdistribution}\label{sec:cond-viable-superd}
At first sight, superdistribution seems a variant of DRM or p2p, or a combination of both.
Now we try to elaborate on specific traits to show how superdistribution is different.
\subsection{The axis of lawfulness and legitimacy}
The RON, if effective, turns superdistribution into a network marketing scheme,
or if multiple buyer generations receive remunerations, a multi-level marketing scheme.
Multi-level marketing carries negative connotations and is illegal 
 in special forms known as pyramid selling, snowball systems, chain-letters,
etc., under many jurisdictions.
This similarity to illicit schemes has perhaps also impeded applied research in
the field of superdistribution as such.
\cite[Vol. II]{Micklitz1999} present criteria to distinguish 
between legitimate multi-level marketing
and such practises that are to be considered illicit.
In the case of digital goods some arguments speak for the viability of fair
superdistribution schemes (thoroughly discussed in~\cite{AUS05A,AUS04C}).
i) Buyers acquire not
only a void right to resale, but also a good of value.
Potential losses an agent entering at a late stage will incur 
are charged up against this value;  
ii) \textit{Inventory loading}\ie the obligation to keep a large, non-returnable stock,
is irrelevant for digital goods;
iii)  Marginal costs for replication and redistribution
are mostly much smaller than resale prices
and thus transaction costs are largely insignificant;
iv) A main novel feature of the concepts above is that they enable
in principle a fair system design (see below).

Other legal requirements for superdistribution are derived from the corresponding
ones for general electronic commerce.
i) privacy of buyers and sellers should be maintained by implementing 
minimal-need-to-know principles; 
ii) Consumer protection legislation, as\eg in the EU~\cite{EUcons}, needs
to be respected;
iii) Copyright law must be respected\ie originators rights must
be properly transcribed into the licences and a system's operator
must obtain all necessary rights and involve collecting societies, etc.
iv) Contracts between buyers and resellers must be enforceable and individual
fraud (e.g., by selling content of lower value than proposed) must be prevented;
v) Market abuse and distortion must be prevented, cf.~the economical and
security requirements below.
\subsection{The economical axis}\label{sec:economical-axis}
Digital goods share the properties of information goods which are transferable and
non-rival like public goods, and additionally are durable\ie show no
wear out by usage or time~\cite{b:SV99,STEG04}.  
Like for a private good, however, original creation can be costly, 
whereas reproduction and redistribution are potentially very cheap.
This is the economical basis for superdistribution which emulates 
the distribution system of 
free-riders, namely p2p networks~\cite{ZK04}.
They pose additional value proposition  to buyers of the
original (legal) version of the good by revenues or other rewards 
linked to resales.
Thus the central question for superdistribution of digital goods is 
economic viability in the presence of free-riders.

\begin{figure}[h]
\centerline{\resizebox{0.22\textwidth}{!}{\includegraphics{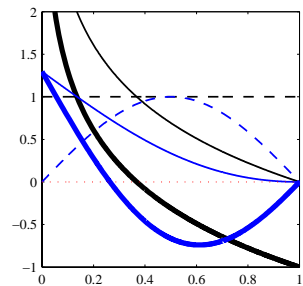}}} 
\caption{Examples for expected revenues from resales and effective price in random superdistribution
with dynamical forward pricing.}
\label{fig:incentive_examples}
\end{figure}

The RON of a superdistribution network is a network marketing scheme.
Theoretical treatments for network markets are scarce, which inspired us
to devise a stochastic model for the dynamics of such markets in~\cite{AUS05A}
and evaluate it in various ways~\cite{AUS04C,AUS_HICSS2008}.
The model is essentially comprised of atomic agents entering the market
\textit{continuously} until saturation, with equal chance to trade
with each other\ie to buy the good from a reseller.
With these assumptions, a node entering the CDO at a certain point in time\ie
a certain market saturation, can calculate its \textit{expected} revenues from
subsequent resales, given that the \textit{price schedule} of current and future
resales prices is known.
Figure~\ref{fig:incentive_examples} shows two examples (black, blue) of prices
(dashed), expected resales revenues (thin solid) and effective prices\ie 
price paid minus expected revenues (thick solid), plotted against the
saturation parameter running from $0$ to $1$.
The thrilling flip side of the innocuous mathematical expressions defining this
model is that it enables \textit{dynamical forward pricing}.
That is, the operator of a superdistribution network can in principle
control the incentive that accrues to buyers via the resales revenues over time.
This possibility has not been exploited by any superdistribution
schemes yet.

Further results  model's analysis spark optimism for superdistribution
as a business and its viability as a replacement for DRM. 
In a basic extension of the model it was shown in~\cite{AUS_HICSS2008} that the legitimate good
in the CDO can prevail against a free-rider version under moderate assumptions.
Nonetheless, superdistribution market mechanisms need to be carefully crafted
as many more external factors other than rational decision-making based on pecuniary 
incentives come into play.
One important aspect in that vein is  market homogeneity.
While superdistribution will work fine in a population which
consists of a rather homogeneous group of individuals,
for instance with special preferences, it may break down if the market
is biased in the sense that there is a group of agents with
higher trading capacities\eg large
music labels running direct sale web sites.
Furthermore, inhomogeneities amplified by network effects~\cite{ECON96b,ECON96a,MH03,SWA02,LCP03}
carry the imminent danger that the market can be cannibalised
at an early stage by an agent with overwhelmingly high
communication capacity\eg a popular web site, who could
then obtain a practical monopoly. 

Finally, there is a psychological element to superdistribution
that is connected to the aleatory element of network markets and 
human sense of justice, which modern empirical economics has shown to be 
an important driving force of human action~\cite{Fehr2000}.
In the small-scale study on a real superdistribution system~\cite{AHPB08},
it was shown that users felt bad about the monetary incentive they received
from resales since they were asking money from their peers for something
that was perceived as pure entertainment.
Though these results may be culture-dependent to some extent, 
they show that the marketing aspects of superdistribution deserve 
utmost care.
\subsection{The security and technical axis}\label{sec:techn-secur-axis}
From a security viewpoint the central difference between DRM and superdistribution is that
DRM protection is focused entirely on the CDO, while in superdistribution the most 
important protection goals regard the remuneration.
In fact, the parts of superdistribution which require local DRM protection in and
between the nodes are encoded in the consumption rules of the consumption licence and the 
redistribution rules.
The latter are essential to protect the business model and market mechanism implemented
by the superdistribution system's operator.
These CDO protection requirements can be implemented by arbitrary DRM measures, centralised
or de-centralised and with a varied level of enforcement, 
as we have seen in the examples of Section~\ref{sec:two-examples}.
An important point for the buyer is the secure association to the content
to prevent the mentioned content masquerading.
On the other hand, ensuring remuneration is essential to implement a fair 
superdistribution market.
A natural way to combine the in-band with the necessary out-of-band processes\eg payment,
is by sending back receipts, which are cryptographically bound to the content and transaction,
to the reseller.
The reseller can then for instance redeem these receipts as tickets at a central rewarding service.
We have shown a way to implement such general schemes with trusted platforms in~\cite{KuntzeSchmidt2007B}.

Privacy is of utmost importance in a network of transactions involving a large number of partners.
In superdistribution privacy is limited again essentially in the remuneration process,
since there buyer and reseller need to reveal their identities and transaction data toward a
payment provider or transaction processing service.
This is not a gross risk to privacy, since often buyers and resellers are acquainted anyway,
for instance if superdistribution is based on personal recommendation.
In general, the identities of nodes in the RON should be protected by Identity
Management systems~\cite{Clau2001,Pfitzmann2005} to the appropriate level.
The Paradiso system described in Section~\ref{sec:drm-paradiso} exhibits the
usual trade-off between security and privacy.
The chain of licences transported downstream in the CDO contains information
(though not necessarily personalised) on every superdistribution transaction on a path.
It would be interesting to see if security can be protected with similar strength
but with higher privacy levels.
Methods for that can for instance make use of cryptographic zero-knowledge proofs~\cite{Feige1998,Camenisch2002}. 

Superdistribution as such is almost technology-neutral.
Three challenges need to be met for their success in the economy of digital goods:

\textit{Market mechanisms} must be implementable in a general superdistribution
framework or platform. Such a framework should enable the definition of CDO
and RON structures, for instance rewarding levels, match-making rules,
allowed number of resales, or the more concrete rules some of which have
been mentioned in Section~\ref{sec:examples}.

A \textit{marketing platform} must be incorporated in the superdistribution
network, in particular to ensure fairness in trade and competition between resellers,
and market homogeneity. 

The \textit{dynamisation} of the market should be supported.
This regards local changes in space and/or time of the two licences,
of which perhaps the most important example is dynamical forward pricing.
A related research challenge is to devise methods to monitor the market in real time.
This would for instance be useful to furnish up-to-date information on the popularity
of a piece of content.

As an example, the digital good could be made returnable to the
originator or the reseller if the chances to achieve further resales revenues
becomes to low.
\section{Some reflections on copyright protection, user-generated content, and free riding}\label{sec:some-refl-copyr}
What is the relationship between superdistribution and copyright protection?
Most existing superdistribution systems use copyright protection on the content, thus raising all
well-known questions of fair use. In those systems, superdistribution is just another
marketing scheme for copyright-protected content. One exception is the Potato system which
deliberately refrains from applying copyright protection on the content ans therefore
represents a true alternative to rigid DRM.
As was shown theoretically in previous work, DRM-free superdistribution can in fact be economically viable
even in the presence of free riders due to the incentives provided to legitimate resellers in the
network marketing scheme implemented via the remuneration overlay, cf.~Section~\ref{sec:economical-axis}.
Thus, the data that needs the strongest protection in superdistribution is the redistribution licence,
not the consumption licence nor the content itself, as we found out in Section~\ref{sec:techn-secur-axis}.

Nevertheless, technical copyright protection is the prevalent method used in the marketing of 
digital goods, and centralised distribution is dominant. So it is interesting to reflect
on the ongoing ``battle'' between copyright holders and ``pirates'' to see what role superdistribution
of digital content may play in the future.
So, free-riders are fought by technical methods for copyright protection
and accompanying legal regulation~\cite{DMCA,WCT,EU-CopyrightDirective},
which, generally speaking, aim at restoring features of
private, physical goods. 
None of this has lead to sustainable success and economic, legal, and societal 
implications of rigid DRM raised a heated debate
about its various
fundamental, economic, and pragmatic
problems~\cite{DRM}, cf.~\cite{IEP-SI04} for a more general
discussion of the underlying concepts of intellectual property rights.
The general legitimacy of DRM measures which tend to disrupt
consumers' expectations on their individual usage of the
good~\cite{MHB03}, is doubtful in light of empirical findings on the
effect of illegal file-sharing on record sales~\cite{OS04}, which
seems negligible.

Some industry players ``defect from the front'', for instance
iTunes now offers media from major label EMI
with superior quality and free 
of copy protection,
using the absence of DRM as a
means of quality discrimination~\cite{FT-apple-copyright-free}.
Exponents of the computer and media industries issued statements raising doubts
on the viability of DRM for media marketing~\cite{jobs-thoughts-on-music,betanews-jupiter-DRM-study}.
On the other hand recent court cases had a mixed outcome for both sides, sometimes awarding
(punitive) damages to copyright holders, sometimes questioning the legitimacy of the case as such.
Nevertheless  the legal lever of copyright holders is becoming
unprecedentedly long. In an unfortunate turn of affairs of historic dimension this is
associated with novel legislation on lawful interception and mandatory data retention
in telecommunication meant to protect societies from serious organised crime and terrorism.
The pressure of the industries' lobbyists is now on ISPs to filter copyrighted content and block
users even on the mere suspicion of infringement.
This approach has failed in the first attempt to push it through at the European Parliament's
cultural commission~\cite{EFF08}, 
and the Commissioner for the Internal Market has rejected the implementation of EU policies
 in that direction~\cite{CMC08}.
Nonetheless, France has enacted legislation to bar 
users from the Internet as penalty for copyright violations~\cite{IFPI-France}.
The demands go so far as to ``outlaw'' and completely filter p2p protocols~\cite{EFF07},
as they are mostly used for ``piracy''\footnote{With this rationale, SMTP should be outlawed,
as the overwhelming proportion of e-mail is SPAM.}. 

As Lawrence Lessig states in his insightful talk~\cite{lessig-talk}, ``there is growing extremism 
that comes from both sides in this debate [\ldots]''. On the one side, a abolitionist 
attitude toward new technology which for instance automatically removes copyrighted content
from sites like YouTube, regardless whether  there might be a claim of fair use to it or not.
On the other side a growing disrespect among the youth for the concept of intellectual property
as such, and even for the law in general. User-generated content relying on original content, such as 
``remixes'' found frequently on the Web are a good example of new forms of creativity that
may be thwarted in such a hostile environment. The implications go beyond digital goods and could 
impact the whole way the Internet is used, as every move in it by a law-abiding citizen or his 
children bears incalculable risk of being incriminated. In such a situation there is in fact little
reason for copyright holders to experiment with alternatives to centralised content distribution
protected by DRM, civil, and penal law. Culture would be ``read-only'' as Lessig phrases it.
Making the historical analogy to the advent of broadcasting technology in the US in the 1930s,
Lessig describes how the then ruling cartel ASCAP that controlled most of the performance rights
nearly strangled the new media by charging broadcasters inflationary prices. This worked until in 1939
Broadcast Music Incorporated (BMI) was founded. BMI was a content aggregator organised much more 
democratically and providing its subscribers for instance with bundles of musical works from the 
public domain at economic prices. In the early 1940s most broadcasters switched to BMI.
ASCAP countered with content quality arguments that are resounding in the current DRM debate
as well. Nevertheless ASCAP cracked in 1941 and the bottom line  is that competition
alone was enough to break a legal cartel over access to content.
Thus emerges the strongest argument against the advent of the read-only culture.
There can be an economic balance 
between copyright holders and consumers and it can be struck by the counterweight represented 
by community-produced, user-generated content such as remixes and mashups.

In view of all this and with hindsight to history, superdistribution and particularly network markets
for digital goods could be a part of the economic counterweight necessary to strike the mentioned
balance. As the Web is evolving now to the Web~2.0 where user-generated content and communities
gain an increasing importance, two germ ideas could be followed that might raise the economic impact
of superdistribution. First, an open superdistribution platform can be envisaged on which everyone
can set up his own market mechanism for his own content. Second, in a given superdistribution
system, resellers could be explicitly allowed and even encouraged to create derivative works
of the original content they superdistribute to become value-added resellers.
Both ideas enable resellers in a CDN to differentiate themselves from each other. This
helps to provide an equal opportunity market for all participants and make it more homogeneous.
\section{Conclusion}\label{sec:conclusion}
The main claim of the present paper is that superdistribution is conceptually different
from both DRM and p2p and is really a third field in its own right.
In fact we have shown that the system theoretic content of superdistribution
is much richer than that of DRM systems, as it uses for the first time --- and by necessity ---
informational representations for the value proposition of a digital good to its 
buyers\ie the combination of consumption and remuneration for resales.
Moreover the economy of superdistribution lies on a categorically different
level than the economy of p2p networks, which is centred on questions of 
incentives for participation and fairness in the contribution of resources~\cite{ACM04},
rather than transported values.

We conclude that the evolution of superdistribution based business models for digital goods
is still in its early beginnings --- and though the risks are considerable, the prospects
are equally thrilling.
As a research subject, superdistribution can be really attractive since it is interdisciplinary
by nature and at the same time has a clearly defined field of experiment in the digital economy.







%



\providecommand{\noopsort}[1]{} \providecommand{\singleletter}[1]{#1}

\end{document}